\documentstyle[12pt, axodraw]{article}

\newcommand{\bp}{{1\over 2}\sqrt{y^4_R+y^4_I} + {1\over 2}y^2_R}
\newcommand{\bm}{{1\over 2}\sqrt{y^4_R+y^4_I} - {1\over 2}y^2_R}
\newcommand{\cp}{{1\over 2\sqrt{y^4_R+y^4_I}} + {y^2_R\over 2(y^4_R+y^4_I)}}
\newcommand{\cm}{{1\over 2\sqrt{y^4_R+y^4_I}} - {y^2_R\over 2(y^4_R+y^4_I)}}
\newcommand{\lo}{{\rm ln}\Big(y^4_R+y^4_I\Big)}
\newcommand{\ta}{{\rm tan}^{-1}\Big({y^2_I\over y^2_R}\Big)}
\newcommand{\gh}{{\hat g}}
\newcommand{\hf}{{1\over 2}}
\begin{document}

\begin{center}
{\Large\bf Gap equation in scalar field theory at finite temperature}\\[6mm]
Krishnendu Mukherjee \footnote{Electronic address: 
krish@tnp.saha.ernet.in}\\[4mm]
{\sl Saha Institute of Nuclear Physics, 1/AF Bidhannagar, 
Calcutta- 700 064, India}
\end{center}
\vspace{1.5cm}

\begin{abstract}
We investigate the two-loop gap equation for the thermal mass of hot massless
$g^2\phi^4$ theory and find that the gap equation itself has a non-zero
finite imaginary part. 
This indicates that it is not possible
to find the real thermal mass as a solution of the gap equation beyond
$g^2$ order in perturbation theory. We have solved the gap equation and 
obtain the real and the imaginary part of the thermal mass which are correct
up to $g^4$ order in perturbation theory.

\vspace{.5cm}

\end{abstract}  

\newpage

It is  well-known \cite{Lind} that if a  theory contains massless bosonic field
such as, QCD or massless scalar theory with $g^2\phi^4$ interaction, then 
at very high temperature $(T)$ the perturbative computations beyond certain order of
coupling constant are afflicted with infrared (IR) singularities. 
In the case of massless $g^2\phi^4$ theory, the one loop contribution to two point
function shows that the fields are screened and the screening mass (Debye mass) 
is found to be of order $gT$ \cite{Alt2,Mat}. However, the result of two loop 
corrections to it is found to be IR divergent. A natural way of avoiding 
this IR divergences in the two loop computation is to use the dynamically 
generated one loop thermal mass $gT$ as the lower cut-off of the momentum
integration. As a result one finds the appearance of a new $g^3$ order 
correction to two point function which, although consistent with the spirit of 
perturbation theory, has not been predicted from the usual perturbative 
expansion in powers of coupling constant $(g)$ at zero temperature.
In addition to that there are infinite number of higher order diagrams that 
contribute to this particular $g^3$ order which in turn is the signature 
of the break-down of usual perturbation theory at very high temperature.
Moreover, it also suggests that one has to resum this infinite number of
diagrams to correctly calculate this $g^3$ order contribution 
\cite{Alt2,Alt4,Wien} to two-point function.

However remaining within the framework of perturbation theory one can in
principle be able to calculate the thermal mass to any order of coupling 
constant by solving the gap equation \cite{Ban,Jack}. In order to obtain the 
gap equation, the functional integral formulation may be used, and we shall 
briefly discuss this method following Jackiw {\it et. al.} \cite{Jack}. 
Consider a massless $g^2\phi^4$ theory in 
$d$ dimension $(d=4-2\epsilon)$ described by the lagrangian
\begin{equation}
{\cal L} = {1\over 2}\partial_{\mu}\phi\partial^{\mu}\phi
-\mu^{2\epsilon}{g^2\over 4!}\phi^4~,
\end{equation}
and the partition function is given by
\begin{equation}
Z = \int {\cal D}\phi  e^{iS(\phi)}~,
\end{equation}
where $S(\phi)=\int{d^d}x{\cal L}(\phi(x))$.
We introduce a loop counting parameter $l$ and write down $Z$ as
\begin{equation}
Z = \int{\cal D}\phi  e^{{i\over l}S(\sqrt{l}\phi)}~.
\end{equation}
In the usual perturbation theory we separate the quadratic part of $S(\phi)$
and expand the exponential of the remainder in powers of $l$. To obtain a gap
equation for a possible mass $m$, we add and subtract $S_m = -{m^2\over 2}
\int{d^d}x \phi^2{(x)}$, which of course changes nothing (at least in the 
classical level, the equation of motion remains the same).
\begin{equation}
S = S + S_m - S_m
\end{equation}
We recognize the loop expansion by expanding $S + S_m$ in the usual way,
but taking $-S_m$ as contributing at one loop higher. These can be 
accounted systematically by replacing Eq. (4) with an effective action $S_l$.
\begin{equation}
S_l = {1\over l}[S(\sqrt{l}\phi) + S_m{(\sqrt{l}\phi)} 
- S_m{(\sqrt{l}\phi)}] ~.
\end{equation}
Starting from this effective action the self energy $\Sigma$ of the complete 
propagator can be calculated to any order in $l$ and set $l=1$ at the end of 
the calculation. The gap equation is obtained by demanding that $\Sigma$ does
not shift the mass $m$, i.e., 
\begin{equation}
\Sigma{(p)}\mid_{p^2 = m^2} = 0
\end{equation}
Furthermore, in order to get a real solution of $m$ from Eq. (6) one has to
ensure that the imaginary part of $\Sigma{(p)}$ at $p^2 = m^2$ is zero.

We apply the above mentioned ideas to $g^2\phi^4$ theory (described by
Eq. (1)) at finite temperature using real time formalism. Let us recall
that in the real time formalism \cite{Nie,Land} the thermal propagator has a 
$2\times{2}$ matrix structure, the $1-1$ component of which refers to the 
physical field, the $2-2$ component to the corresponding ghost field, with the 
off-diagonal $1-2$ and $2-1$ components mixing them. The propagator 
used here is given by
\begin{equation}
\left( \begin{array}{cc}{D_{11}(K)}&{D_{12}(K)}\\ {D_{21}(K)} &{D_{22}(K)}
\end{array}\right) = \left( \begin{array}{cc}
{\Delta_0(K) + \Delta_{\beta}(K)}
&{{\bar\Delta}_{\beta}(K)}\\ {{\bar\Delta}_{\beta}(K)} &{\Delta^*_0(K) + 
\Delta_{\beta}(K)}\end{array} \right)~,
\end{equation}
where $\Delta_0 (K)$ is the usual Feynman propagator at zero temperature
\begin{equation}
\Delta_0(K) = {i\over K^2 - m^2 + i\epsilon}~.
\end{equation}
Here $\Delta_{\beta}$ and ${\bar \Delta}_{\beta}$ are finite temperature 
corrections to the zero temperature propagator where
\begin{equation}
\Delta_{\beta}(K) = 2\pi \delta(K^2-m^2)n_B(|K_0|),~~ {\bar\Delta}_{\beta}(K) = 2\pi \delta(K^2-m^2){\rm exp} 
\left( { \beta|K_0|\over 2} \right) n_B(|K_0|),
\end{equation}
with the Bose-Einstein factor $n_B(K_0) = {1/(e^{\beta|K_0|} -1)}$ $(\beta =1/T)$.
The complete self energy is given by the expression \cite{Land,Kob}:
\begin{equation}
{\rm Re} \Sigma(p) = {\rm Re}\Sigma_{11}(p),~~~ {\rm Im} \Sigma(p) = {i\over 2}
{{e^{\beta p_0}-1}\over e^{\beta p_0/2}} \Sigma_{12}(p),
\end{equation}
where $\Sigma_{11}$ and $\Sigma_{12}$ are the self energy of the $1-1$ component
and $1-2$ component, respectively.

The graphs that are contributing to $\Sigma(p)$ up to two-loop order are 
depicted in Fig. 1. The contributions to ${\rm Re}\Sigma(p)$ coming 
from these graphs, although in a different context, have already been 
computed in Refs. \cite{Alt2,Mat,Ban,Fuj}. We find that the real part of self 
energy in the limit $p_0=m$ and ${\bf p} \to 0$ is given by
\begin{eqnarray}
{\rm Re}\Sigma(p_0=m,{\bf p} \to 0) &=& -y^2 T^2 + {\hat g}^2 T^2\Bigg\{-\pi y 
+ y^2\sum_{n=1}^{\infty}B_n(y) - \sum_{n=1}^{\infty}A_n(y)\Bigg\} \nonumber \\
& &+ {\hat g}^4 T^2\Bigg\{ 2\pi^2 + a_1y - 2\pi y \; {\rm ln} \;y^2 - a_2 y^2 
+ a_3 y^2 \; {\rm ln} \; y^2 \nonumber \\
& &~~~~~~~ - {y^2\over 4}({\rm ln} \; y^2)^2 - (a_4+ {\rm ln} \; y^2-{\pi\over y})
\sum_{n=1}^{\infty}A_n(y) \nonumber \\
& &~~~~~~~ - (a_5y^2-2\pi y)\sum_{n=1}^{\infty}B_n(y) + \sum_{m=1}^{\infty}
\sum_{n=1}^{\infty}A_m(y)B_n(y) \nonumber \\
& &~~~~~~~ + 4 [X(y)+I(y)] \Bigg\} + 0({\hat g}^5)~,
\end{eqnarray}
where $y={m\over T}$, ${\hat g}={g\over 4\pi}$, 
$a_1=15.64-21.99\; {\rm ln} \; ({T\over \mu})$, $a_2=26.07-3.46 
\; {\rm ln} \;({T\over \mu})
-2({\rm ln} \; ({T\over \mu}))^2-{1\over 6}\delta_1 +{1\over 2}\delta_2$,  
$a_3=1.62+{1\over 3} \; {\rm ln} \; ({T\over \mu})$, $a_4=2.54$, 
$a_5=0.98+ {\rm ln} \; ({T\over\mu})$, 
$A_n(y)={1\over n}[8\pi^2 n^2(1+{y^2\over 4\pi^2n^2})^{1\over 2}-y^2]$,
$B_n(y)={1\over n}[(1+{y^2\over 4\pi^2n^2})^{-{1\over 2}}-1]$,

\noindent $ \delta_1=\int_0^1 dx_1\int_0^1 dx_2 (1-x_2)
\; {\rm ln} \; [(1-x_2+{x_2\over {(1-x_1)x_1}}) -x_2(1-x_2)]$,

\noindent $\delta_2=\int_0^1 dx_1\int_0^1 dx_2 ({1\over x_2}-1)
\; {\rm ln} \; [(1-x_2+{x_2\over {(1-x_1)x_1}})-x_2(1-x_2)]$,

\noindent $X(y)=y^2\int_1^{\infty}{dt\over (e^{ty}-1)}t \; {\rm ln}
\; (t+\sqrt{t^2-1})$,

\noindent $I(y)=y^2\int_1^{\infty}{dt_2\over (e^{yt_2}-1)} {\rm PP}
\int_1^{\infty}{dt_1\over (e^{yt_1}-1)} \; {\rm ln} \; 
[{\sqrt{t_2^2-1}-\sqrt{t_1^2-1} \over\sqrt{t_2^2-1}+\sqrt{t_1^2-1}}]$.

The only graph that will contribute to the ${\rm Im}\Sigma(p)$ is 
Fig. 1(h) and its contribution to $\Sigma_{12}(p)$ is given as
\begin{eqnarray}
-i\Sigma_{12}(p) &=& {g^4\over 6}
 \int {d^4K_1\over (2\pi)^4}{d^4K_2\over (2\pi)^4} \left[ 
{\bar \Delta}_{\beta}(K_1) {\bar \Delta}_{\beta}(K_2) 
{\bar \Delta}_{\beta}(p - K_1 -  K_2) \right]  \nonumber \\
&=& {g^4\over 6(2\pi)^5} \int d^3k_1 \; d^3k_2 \; 
{1\over 4E_{k_1}E_{k_2}} n_B(E_{k_1}) n_B(E_{k_2}) \nonumber \\
& & n_B(E_{p-k_1-k_2}) e^{\beta E_{k_1}/2} e^{\beta E_{k_2}/2} 
e^{\beta E_{p - k_1 - k_2}/2} \nonumber \\
&\times & \Bigg\{ \delta\left[ (p_0 - E_{k_1} - E_{k_2})^2 -
E^2_{p - k_1 -k_2} \right] \nonumber \\
& &~ + \delta\left[ (p_0 - E_{k_1} + E_{k_2})^2 - E^2_{p - k_1 -k_2}
 \right] \nonumber\\
& &~ + \delta\left[ (p_0 + E_{k_1} - E_{k_2})^2 - E^2_{p - k_1 -k_2}
 \right]\nonumber \\
& &~ + \delta\left[ (p_0 + E_{k_1} + E_{k_2})^2 -
E^2_{p - k_1 -k_2} \right] \Bigg\}~,
\end{eqnarray}
where $E_{k_1} = \sqrt{{{\bf k}_1}^2 + m^2}$, 
 $E_{k_2} = \sqrt{{{\bf k}_2}^2 + m^2}$, 
 and $E_{p-k_1-k_2} = \sqrt{ ({\bf p} -{\bf k}_1 -{\bf k}_2)^2 + m^2}$. 
In the $p_0=m$ and ${\bf p} \to 0$ limit, the expression for 
$-i\Sigma_{12}(p)$, becomes,

\begin{eqnarray}
-i\Sigma_{12}(p_0=m,{\bf p})&=& {g^4\over {6(2\pi)^5}}\int d^3k_1 d^3k_2 
{1\over E_{k_1}E_{k_2}}n_B(E_{k_1})n_B(E_{k_2}) \nonumber\\
& &~~~ n_B(E_{k_1+k_2})e^{\beta E_{k_1}/2}e^{\beta E_{k_2}/2}
e^{\beta E_{k_1+k_2}/2}\nonumber\\
& &\times\Bigg\{\delta[(m-E_{k_1}-E_{k_2})^2 - E^2_{k_1+k_2}]\nonumber\\
& &~~~+\delta[(m-E_{k1}+E_{k_2})^2 - E^2_{k_1+k_2}]\nonumber\\
& &~~~+\delta[(m+E_{k1}-E_{k_2})^2 - E^2_{k_1+k_2}]\nonumber\\
& &~~~+\delta[(m+E_{k1}+E_{k_2})^2 - E^2_{k_1+k_2}]\Bigg\}\nonumber\\
\end{eqnarray}   

If ${\bf k_1}$ makes an angle $\theta$ with the $k_{2z}$ direction 
then, the above expression becomes

\begin{eqnarray}
-i\Sigma_{12}(p_0=m,{\bf p}\rightarrow 0)&=&{g^4\over 48(2\pi)^4}
\int {d^3k_1\over k_1E_{k_1}}\int_0^\infty{k_2dk_2\over E_{k_2}}
\int_{-1}^{+1}dx n_B(E_{k_1})n_B(E_{k_2})\nonumber\\
& & n_B(E_{k_1+k_2})e^{\beta E_{k_1}/2}e^{\beta E_{k_2}/2}\nonumber\\
&\times&\Bigg\{\delta(x-x_1)\theta(1-|x_1|)+\delta(x-x_2)
\theta(1-|x_2|)\nonumber\\
& &~~+\delta(x-x_3)\theta(1-|x_3|)+\delta(x-x_4)\theta(1-|x_4|)
\Bigg\}
\end{eqnarray}    

where 

\begin{eqnarray}
x &=& cos\theta \nonumber\\
x_1&=&{{(m-E_{k_1}-E_{k_2})^2-{\bf k_1}^2-{\bf k_2}^2-m^2}\over 
2k_1k_2};x_2 = {{(m-E_{k_1}+E_{k_2})^2-{\bf k_1}^2-{\bf k_2}^2-m^2}\over 
2k_1k_2} \nonumber\\
x_3&=&{{(m+E_{k_1}-E_{k_2})^2-{\bf k_1}^2-{\bf k_2}^2-m^2}\over 
2k_1k_2};x_4 = {{(m+E_{k_1}+E_{k_2})^2-{\bf k_1}^2-{\bf k_2}^2-m^2}\over 
2k_1k_2}
\end{eqnarray}  

Now we have to evaluate the range of integrations over ${\bf k_1}$ and
${\bf k_2}$ using this theta functions.

$\quad 1)$ $\theta(1-|x_1|)=\theta(1-x_1)\theta(1+x_1)$. Therefore $x_1\leq 1$
implies 
\[ \sqrt{E_{k_1}-m}\sqrt{E_{k_2}-m}\Bigl(\sqrt{(E_{k_1}+m)(E_{k_2}+m)} 
-\sqrt{(E_{k_1}-m)(E_{k_2}-m)}  \Bigr)\geq 0 ,\]
and it is trivially satisfied for positive values of $k_1$ and $k_2$.
Similarly $x_1\geq -1$ implies 
\[(E_{k_1}-m)(E_{k_2}-m)+k_1k_2\geq 0 ,\]
and is also satisfied by both of $k_1$ and $k_2$. Therefore, this
theta function does not put any extra constraint on the integration 
range of $k_1$ and $k_2$.  

$\quad 2)$ $\theta(1-|x_2|)=\theta(1-x_2)\theta(1+x_2)$. $x_2\leq 1$
implies \[(E_{k_1}-m)(E_{k_2}-m)+k_1k_2\geq 0 ,\] and it is trivially
satisfied for positive values of both of $k_1$ and $k_2$. 
Next $x_2\geq -1$ implies \[k_2(k_2-k_1)\geq 0 .\] Therefore it gives
$\theta(1-|x_2|)=\theta(k_2-k_1)$.

$\quad 3)$ $\theta(1-|x_3|)=\theta(1-x_3)\theta(1+x_3)$. Proceeding in 
the similar fashion one gets $\theta(1-|x_3|)=\theta(k_1-k_2)$.

$\quad 4)$ $\theta(1-|x_4|)=\theta(1-x_4)\theta(1+x_4)$. $x_4\leq 1$
implies \[k_2(k_2+k_1)\leq 0 ,\] and this relation is not been satisfied
for positive values of $k_1$ and $k_2$. Therefore this theta function 
will not give any contribution to $-i\Sigma_{12}(p_0=m,{\bf p}\rightarrow
0)$.

Using this theta function constraints we ultimately get, 

\begin{eqnarray}
-i\Sigma_{12}(p_0=m,{\bf p}\rightarrow 0) &=& {g^4\over 24(2\pi)^3}
\int_0^\infty {dk_1k_2\over E_{k_1}}\int_0^\infty {dk_2k_2\over E_{k_2}}\nonumber\\
& & n_B(E_{k_1}) n_B(E_{k_2})e^{\beta E_{k_1}/2} 
e^{\beta E_{k_2}/2}\nonumber\\
&\times&\Bigg\{ n_B(E_{k_1}+E_{k_2}-m)
e^{\beta(E_{k_1}+E_{k_2}-m)/2}\nonumber\\    
& &~~~+\theta(k_2-k_1)n_B(E_{k_2}-E_{k_1}+m)
e^{\beta(E_{k_2}-E_{k_1}+m)/2}\nonumber\\
& &~~~+\theta(k_1-k_2)n_B(E_{k_1}-E_{k_2}+m)
e^{\beta(E_{k_1}-E_{k_2}+m)/2}\Bigg\}
\end{eqnarray}

Now if we interchange $k_1$ and $k_2$ in the last term of the 
above expression it becomes identical to the second term, and
we get,

\begin{eqnarray} 
-i\Sigma_{12}(p_0=m,{\bf p}\rightarrow 0) &=& {g^4\over 24(2\pi)^3}
\Bigg\{ \int_0^\infty {dk_1k_1\over E_{k_1}}\int_0^\infty 
{dk_2k_2\over E_{k_2}}e^{\beta(E_{k_1}+E_{k_2})}e^{\beta m/2}
n_B(E_{k_1}+E_{k_2}-m)\nonumber\\
& &~ + \int_0^\infty {dk_2k_2\over E_{k_2}}\int_0^{k_2} 
{dk_1k_1\over E_{k_1}}e^{\beta E_{k_2}}e^{\beta m/2}
n_B(E_{k_2}-E_{k_1}+m)\Bigg\}\nonumber\\
& &\times n_B(E_{k_1})n_B(E_{k_2})
\end{eqnarray}

Finally the ${\rm Im}\Sigma(p_0=m,{\bf p}\rightarrow 0)$ takes the
following form:

\begin{eqnarray}
{\rm Im}\Sigma(p_0=m,{\bf p}\to 0) &=& {i\over 2}{{e^{\beta m}-1}\over
 e^{\beta m/2}}\Sigma_{12}(p_0=m,{\bf p}\rightarrow 0)\nonumber\\ 
&=& -{2\pi\over 3}{\hat g}^4T^2 J(y)
\end{eqnarray}

where ${\bar n}_B(x)=1/(e^x-1)$ and

\begin{equation}
J(y) = 2\left ({\rm ln}(1-e^{-y})\right )^2 - 3\int_y^\infty dx
 {\rm ln}(1-e^{-x}) - 2\int_y^\infty dx (x-y){\bar n}_B(x+y)
\end{equation}

Therefore it is evident from Eqs. (11) and (18) that  
$\Sigma(p)\mid_(p_0=m,{\bf p} \to 0)$ is complex  
Consequently, the thermal mass($m$) that we may obtain 
by solving the gap equation $\Sigma(p)\mid_(p_0=m,{\bf p}\to 0)=0$ 
has an imaginary part. 
However, it is clear from Eq. (10) that due to a multiplicative factor 
$(e^{\beta p_0} - 1)$, ${\rm Im}\Sigma(p)$ is zero at $p_o =|{\bf p}|$
and $|{\bf p}| \to 0$ limit and hence $\Sigma(p)$ is real in this limit. 

At very high temperature $ y\ll 1$ and the real and imaginary part of 
the self energy in this high temperature limit takes the  following 
form:

\begin{eqnarray}
{\rm Re}\Sigma(y)&=& -T^2 y^2 + {\hat g}^2 T^2 \Big\{{2\pi^2\over 3} - \pi y
\Big\} + {\hat g}^4 T^2 \Big\{3.7\pi^2 - {2\pi^3\over 3}
{1\over y} - b_1 y - b_2 y^2 \nonumber\\
& & + \Big({4\pi^2\over 3}+2\pi\Big){\rm ln}y^2 - b_3 y^2 {\rm ln}y^2\Big\} 
+ 0({\hat g}^5)
\end{eqnarray}

and

\begin{equation}
{\rm Im}\Sigma(y) = {\hat g}^4 T^2\Big\{{2\pi^3\over 9} + 7.9 y -
{\pi\over 2}y^2 + {\pi\over 3}y{\rm ln}y^2 - {\pi\over 3}({\rm ln}y^2)^2
\Big\} + 0({\hat g}^5)
\end{equation}

where, $b_1= 37.12 - a_1$, $b_2= a_2 - 7.36$ and $b_3= 2.7 - a_3$. Therefore 
the gap equation in this limit is

\begin{equation}
{\rm Re}\Sigma(y) + i{\rm Im}\Sigma(y) = 0.
\end{equation} 

Since $y$ is complex, we set $y^2=y^2_R+iy^2_I$ and substitute it in
the high temperature approximated gap equation(22) to obtain the following
two equations:

\begin{equation}
y^2_R = f_R(y_R, y_I)
\end{equation}

\begin{equation}
y^2_I = f_I(y_R, y_I)
\end{equation}

where $y_R$ and $y_I$ are the real and the imaginary part of the 
thermal mass respectively in the high temperature limit and the
explicit form of $f_R$ and $f_I$ are given in the appendix. The eqn.(23)
and eqn.(24) are transcendental equations in two variables $y_R$ and
$y_I$ and we can solve this equation by the method of iteration\cite{Scar}  
assuming that our solution is correct up to order ${\hat g}^4$. We have 
started first with an approximate values of a pair of roots.

\[ y_R^{(0)^2}={2\pi^2\over 3}{\hat g}^2+3.7\pi^2{\hat g}^4, 
~~~~ y_I^{(0)^2}={2\pi^3\over 9}{\hat g}^4 \] 

After third iteration we find that the improved pair of roots 
are equal to the pair of roots obtained after second iteration.
Therefore at very high temperature the
real and the imaginary part of the thermal mass
up to ${\hat g}^4$ order is

\begin{eqnarray}
m^2_R &=& {2\pi^2\over 3} T^2{\hat g}^2 - \sqrt{{8\over 3}}\pi^2 T^2
{\hat g}^3 + \Big[ 3.7\pi^2 + \Big( {4\pi^2\over 3}+2\pi\Big)
{\rm ln}\Big({2\pi^2\over 3}\Big) \Big] T^2{\hat g}^4\nonumber\\ 
& & - \Big({8\pi^2\over 3}+4\pi\Big) T^2{\hat g}^4 
{\rm ln}\Big({1\over {\hat g}}\Big) + 0({\hat g}^5)
\end{eqnarray}

and

\begin{equation}
m^2_I = \Bigl[ {2\pi^3\over 3} - {\pi\over 3}\Big(
{\rm ln}\Big({2\pi^2\over 3}\Big)\Big)^2\Big] T^2{\hat g}^4
+ {4\pi\over 3} T^2{\rm ln}\Big({1\over {\hat g}}\Big) 
- {4\pi\over 3} T^2\Big( {\rm ln}\Big({1\over {\hat g}}\Big)\Big)^2
+ 0({\hat g}^5)
\end{equation}

We see that up to $g^4$ order both $m^2_R$ and $m^2_I$ are independent of
the ultraviolet scale $\mu$ used in the theory. However from the 
structure of $f_R$ and $f_I$, it is evident that beyond $g^4$ order 
they ought to be $\mu$-dependent.  The result of $m^2_R$ up to order
$g^3$ matches with that obtained by Parwani\cite{Raj}, however 
his $g^4$ order term is $\mu$ dependent. As a result the two
loop real thermal mass obtained in \cite{Raj} becomes unstable
below some characteristics scale of the order of $g^{4/3}T$.

In this letter we have studied the finite temperature gap equation in 
massless $g^2\phi^4$ theory and its nature of solutions. We find that
the thermal mass up to two-loop order which one may obtain 
self-consistently  by solving this 
equation is complex. In the massless $g^2\phi^4$ theory the typical
one-loop thermal mass is of order $gT$ and the expected contribution 
to the mass from the two loop level would be of
order $g^2T$. It is worth mentioning that there may be some non-pertubative
features that really start from this $g^2T$ scale which makes the computation 
unreliable beyond $g^2$ order. It will be also interesting to extend
this method to $3+1$ dimensional QCD where the generation of magnetic 
mass is quite problematic due to IR divergences in the two-loop level.
 
\smallskip  
I gratefully acknowledge Prof. S. Mallik for helpful discussions. I also thank 
the referee for pointing out a major mistake in the evaluation of the
imaginary part of the self energy and also for his suggestion to solve
the gap equation.

\newpage
\renewcommand{\theequation}{A.\arabic{equation}}
\setcounter{equation}{0}
\begin{center}
{\bf Appendix}
\vspace{0.3cm}
\end{center}

\begin{eqnarray}
f_R(y_R, y_I)&=& \gh^2\Big\{{2\pi^2\over 3} - \pi\Big[\bp\Big]^\hf\Big\}
\nonumber\\ 
& &+ \gh^4\Big\{3.7\pi^2 - {2\pi^3\over 3}\Big[\cp\Big]^\hf - 
b_1\Big[\bp\Big]^\hf \nonumber\\
& &- b_2 y^2_R + \Big(\pi + {2\pi^2\over 3}\Big)\lo - \hf b_3 y^2_R\lo 
+ b_3 y^2_I\ta \nonumber\\
&  &- 7.9\Big[\bm\Big]^\hf + {\pi\over 2}y^2_I 
- {\pi\over 6}\Big[\bp\Big]^\hf\lo\nonumber\\
& &- {\pi\over 3}\Big[\bp\Big]^\hf\ta + {\pi\over 3}\ta\lo\Big\}\nonumber\\
& &+ 0(\gh^5)
\end{eqnarray}

\begin{eqnarray}
f_I(y_R, y_I)&=& \gh^2\Big\{-\pi\Big[\bm\Big]^\hf \Big\}\nonumber\\
& &+ \gh^4\Big\{{2\pi^3\over 9} + {2\pi^3\over 3}\Big[\cm\Big]^\hf 
- b_1\Big[\bm\Big]^\hf - b_2 y^2_I\nonumber\\
& &+ \Big(2\pi + {4\pi^2\over 3}\Big)\ta - \hf b_3 y^2_I\lo
- b_3 y^2_R\ta - {\pi\over 2}y^2_R\nonumber\\
& &+ 7.9\Big[\bp\Big]^\hf + {\pi\over 6}\Big[\bp\Big]^\hf\lo\nonumber\\
& &- {\pi\over 3}\Big[\bm\Big]^\hf\ta - {\pi\over 12}\Big(\lo\Big)^2
\nonumber\\
& &+ {\pi\over 3}\Big(\ta\Big)^2\Big\}\nonumber\\
& &+ 0(\gh^5)
\end{eqnarray}

\newpage
{\large{\bf Figure caption}}
\vspace{1cm}

Fig. 1. Graphs contributing to the two point function. Solid cross and solid
dot denote vertices for renormalisation counterterms of $\phi^2$ and
$\phi^4$ type respectively, while a dashed cross represents the 
additional $\phi^2$ insertion.  

\newpage

\vglue 3cm

\begin{center}
\begin{picture}(450,450)(0,0)
\Line(30,400)(130,400)
\CArc(80,414)(14,0,360)
\Text(80,382)[]{(a)}
\Line(180,400)(280,400)
\Line(223,407)(237,393)
\Line(223,393)(237,407)
\Text(230,382)[]{(b)}
\Line(330,400)(430,400)
\DashLine(373,407)(387,393){3}
\DashLine(373,393)(387,407){3}
\Text(380,382)[]{(c)}
\Line(30,300)(130,300)  
\CArc(80,314)(14,0,360)
\CArc(80,342)(14,0,360)
\Text(80,282)[]{(d)}
\Line(180,300)(280,300)
\CArc(230,314)(14,0,360)
\Line(223,335)(237,321)
\Line(223,321)(237,335)
\Text(230,282)[]{(e)}
\Line(330,300)(430,300)
\CArc(380,314)(14,0,360)
\Vertex(380,300){5}
\Text(380,282)[]{(f)}
\Line(30,200)(130,200)
\CArc(80,214)(14,0,360)
\DashLine(73,235)(87,221){3}
\DashLine(73,221)(87,235){3}
\Text(80,178)[]{(g)}
\Line(180,200)(280,200)
\CArc(230,200)(14,0,360)
\Text(230,178)[]{(h)}
 
\end{picture}\\

\end{center}
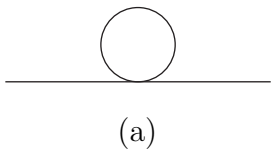
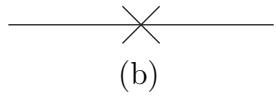
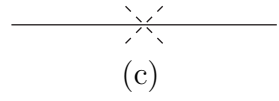
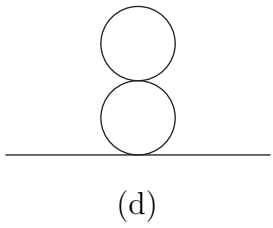
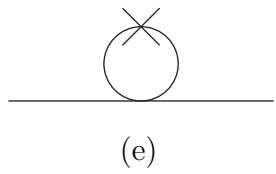
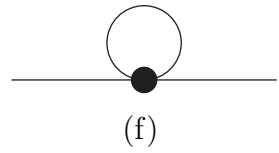
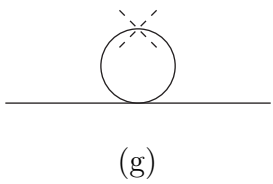
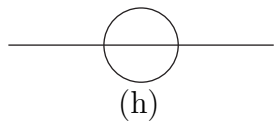
\end{document}